\documentclass[
aps,prd,
twocolumn,
preprintnumbers,
amsmath,
amssymb,
nofootinbib,
preprintnumbers,
superscriptaddress
]{revtex4-1}
\usepackage{here}
\usepackage{footnote}
\usepackage{comment,braket}
\usepackage{epsf}
\usepackage{amsmath}
\usepackage{graphics}
\usepackage{amsfonts}
\usepackage{amssymb}
\usepackage{latexsym}
\usepackage{color}
\usepackage{natbib}
\usepackage{graphicx}
\usepackage{hyperref}
\usepackage{array,multirow}
\usepackage{mathtools}

\usepackage{txfonts}

\usepackage[svgnames]{xcolor}
\definecolor{phthaloblue}{rgb}{0.0, 0.06, 0.54}
\hypersetup{
    colorlinks=true,
    linkcolor=blue,
    citecolor=blue,
    filecolor=blue,
    urlcolor=blue,
    }

\newcommand{\etal}{\textit{et al.\ }}

\begin{document}
\title{Probing Direct Waves in Black Hole Ringdowns}
\author{Naritaka Oshita}
\affiliation{Center for Gravitational Physics, Yukawa Institute for Theoretical Physics,
Kyoto University, Kitashirakawa Oiwakecho, Sakyo-ku, Kyoto 606-8502, Japan}
\affiliation{The Hakubi Center for Advanced Research, Kyoto University,
Yoshida Ushinomiyacho, Sakyo-ku, Kyoto 606-8501, Japan}
\affiliation{RIKEN iTHEMS, Wako, Saitama, 351-0198, Japan}
\author{Sizheng Ma}
\affiliation{Perimeter Institute for Theoretical Physics, Waterloo, ON N2L2Y5, Canada}
\author{Yanbei Chen}
\affiliation{Burke Institute for Theoretical Physics and Theoretical Astrophsyics 350-17,
California Institute of Technology, Pasadena, California 91125, USA}
\author{Huan Yang}
\affiliation{Department of Astronomy, Tsinghua University, Beijing 100084, China}

\preprint{YITP-25-141}
\preprint{RIKEN-iTHEMS-Report-25}

\begin{abstract}
Merger gravitational waves from binary black hole coalescence carry rich information about the underlying spacetime dynamics.
We analyze merger waves from comparable-mass and extreme-mass-ratio binaries, obtained from numerical relativity and black-hole perturbation theory, respectively, and argue that they are dominated by the prompt wave emissions as the black holes collide.
This signal, which we refer to as the \emph{direct wave}, is modulated by the plunging motion and selectively screened by the gravitational potential of the remnant black hole.
The direct wave typically exhibits a time-dependent frequency and decay rate, but for high-spin remnants $(\gtrsim0.7)$ the ergosphere renders it mode-like, with a quasi-stable instantaneous oscillation frequency close to the superradiant frequency. 
We further estimate its detectability in a GW150914-like system and find that the signal-to-noise ratio can exceed $\sim 10$ with the current ground-based detector network. 
Our results therefore identify the direct wave as a robust observable for analyzing black hole ringdowns in current and future gravitational wave events. 
\end{abstract}

\maketitle
%%%%%%%%%%%%%%%%%%%%%%%%%%%%%%%%
\noindent \textbf{\em Introduction.}
%%%%%%%%%%%%%%%%%%%%%%%%%%%%%%%%
In recent years, we have witnessed remarkable progress in understanding the post-merger phase of binary black hole (BBH) coalescence, including the excitation of quasinormal mode (QNM) overtones \cite{Buonanno:2006ui,Giesler:2019uxc,Giesler:2024hcr,Oshita:2021iyn,Oshita:2024wgt,Mitman:2025hgy,Baibhav:2023clw,Finch:2021iip,Zhu:2023mzv,Li:2021wgz,Ma:2021znq}, nonlinearities \cite{Mitman:2022qdl,Cheung:2022rbm,Ma:2022wpv,Khera:2023oyf,Khera:2024bjs,Ma:2024qcv,Zhu:2024rej,Redondo-Yuste:2023seq,Bucciotti:2024zyp,Bourg:2024jme,Bucciotti:2024jrv,Lagos:2024ekd,Yi:2024elj,Bourg:2025lpd,May:2024rrg,Zhu:2024dyl,Ma:2025rnv,Sberna:2021eui}, and late-time tails \cite{Ma:2024hzq,DeAmicis:2024eoy,Islam:2024vro,DeAmicis:2024not,Ma:2024bed,Cardoso:2024jme}. 
These advances not only enable a more precise interpretation of gravitational-wave (GW) events \cite{carullo2019,bustillo2021,Ghosh:2021mrv,Finch:2022ynt, Ma:2023vvr,Ma:2023cwe,Wang:2023ljx,Correia:2023bfn,Capano:2021etf, Siegel:2023lxl} but also promote BH spectroscopy study \cite{Berti:2025hly} into a high-precision regime.

Around the merger, an appealing question arises: \emph{how much of the merger signals reflect remnant BHs' properties, both linearly and nonlinearly?} Giesler \etal \cite{Giesler:2019uxc} proposed that multiple overtones can describe waveforms immediately after the peak. 
This view was later challenged \cite{Baibhav:2023clw}, and more recent analyses with higher accuracy numerical-relativity waveforms indicate that QNM decomposition becomes reliable only some time after the peak \cite{Giesler:2024hcr,Mitman:2025hgy}. 
In other words, waveforms around the peak cannot be captured by a simple mode-based description. Although tentative proposals exist \cite{Mino:2008at,Zimmerman:2011dx,Chavda:2024awq,DeAmicis:2025xuh},
they were formulated in simplified contexts.  

In this {\it Letter}, we investigate merger waves from both comparable-mass BBH simulations and extreme mass-ratio inspirals (EMRIs), and show that they are dominated by the prompt emission as the companions plunge inside the light ring and ergosphere. We refer to this signal component as the \emph{direct wave}. 
A related feature, the so-called ``horizon mode'', was previously studied in \cite{Mino:2008at,Zimmerman:2011dx}, and argued to oscillate at the superradiant frequency with the decay rate determined by the surface gravity. 
Our analysis corrects this picture by (1) showing that the originally proposed horizon mode vanishes due to gravitational screening and (2) associating the major part of direct waves with the early plunging orbit, which is fast-decaying in time.
We further assess the detectability of the direct wave and find that, in a GW150914-like system, its signal-to-noise ratio (SNR) exceeds 10 with the current LIGO-Virgo detector network during the fourth observing run, thereby establishing it as a robust target in recent new events.
Throughout the manuscript, we adopt $c=G=2M=1$, where $M$ is the remnant BH's mass. 
The spin of the BH is denoted by $a$ and the dimensionless spin is $\chi \coloneqq a/M$.
%%%
%%%
\begin{figure}[t]
\centering
\includegraphics[width=0.8\linewidth]{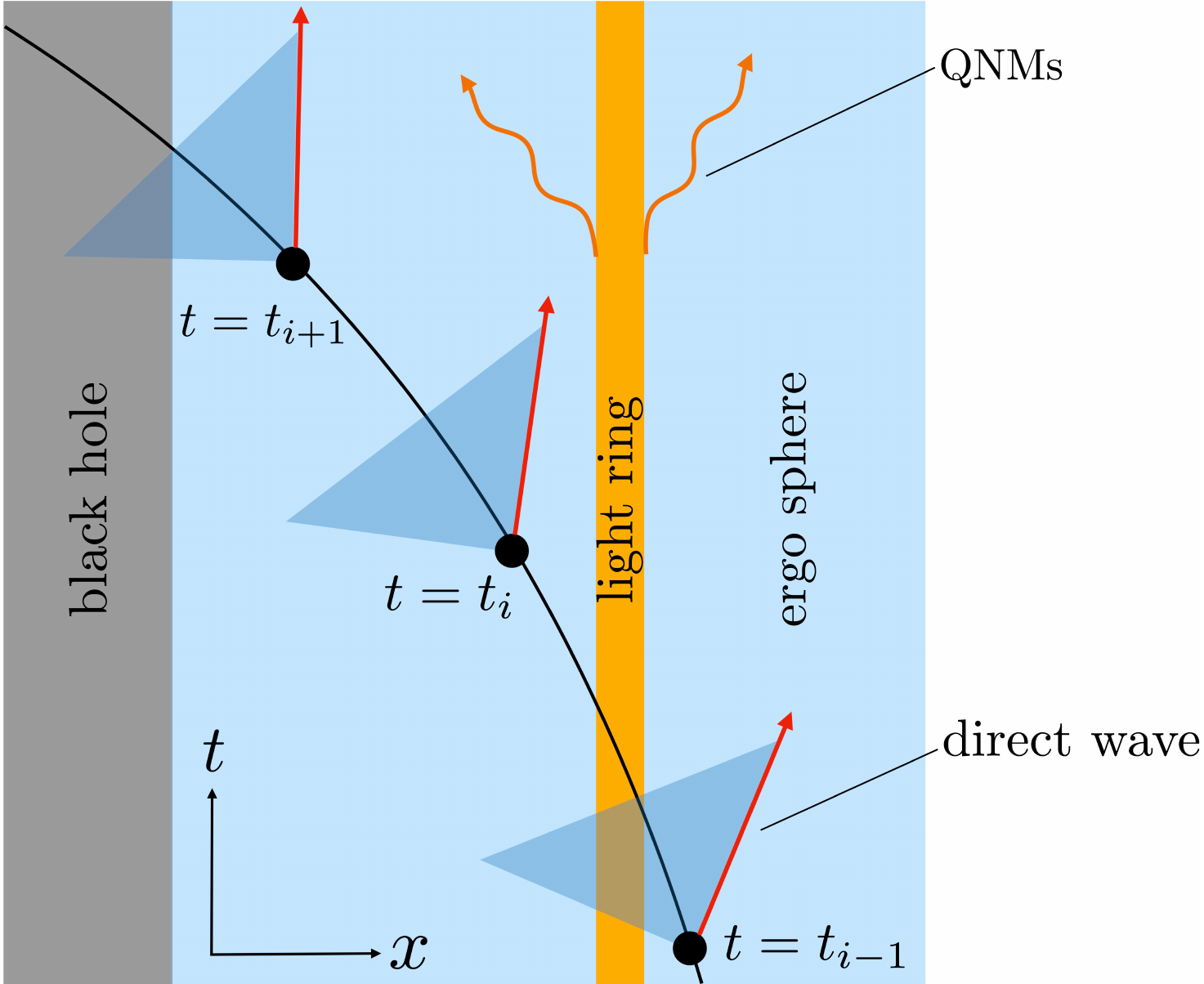}
\caption{
Schematic illustration of a particle plunging into a BH, exciting direct waves (red straight arrows) and QNMs (yellow wavy arrows) nearly simultaneously. 
}
\label{fig_directwaves}
\end{figure}
%%%
%%%

%%%%%%%%%%%%%%%%%%%%%%%%%%%%%%%%
\noindent \textbf{\em Direct waves sourced near BHs.}
%%%%%%%%%%%%%%%%%%%%%%%%%%%%%%%%
GWs from a perturbed BH are described by the Teukolsky equation, with strain at infinity related to the Weyl scalar $\Psi_4$. 
Its spectrum can be decomposed into spheroidal harmonics and a radial Teukolsky variable $R_{\ell m \omega}$, which at infinity reduces to
\begin{align}
\begin{split}
R_{\ell m \omega} &= \frac{r^3 e^{i \omega x}}{2i \omega B_{\ell m \omega}^{\rm in}}
\int_{r_+}^{\infty} dr \frac{R_{\ell m \omega}^{\rm in} {\cal T}_{\ell m \omega}}{\Delta^2(r)} \to r^3 e^{i \omega x} Z_{\ell m \omega}\,,
\end{split}
\label{teukolsky_R}
\end{align}
where ${\cal T}_{\ell m \omega}$ is the source, $r_+$ the outer-horizon radius, and $x$ the tortoise coordinate with $dx/dr=(r^2+a^2)/\Delta(r)$, $\Delta=r^2-r+a^2$. 
The in-mode homogeneous solution $R_{\ell m \omega}^{\rm in}$ behaves as $e^{-i k_{\rm H} x}$ near the horizon, and asymptotes to $B_{\ell m \omega}^{\rm in} e^{-i \omega x}
+ B_{\ell m \omega}^{\rm out} e^{i \omega x}$ at infinity. Here $k_{\rm H} \coloneqq \omega - m \Omega_H$ and the horizon frequency $\Omega_{\rm H}=\chi/(2r_+)$.

When the particle approaches the horizon for $t \geq t_0$, the source simplifies and the spectral function $Z_{\ell m \omega}$ in \eqref{teukolsky_R} reduces to \cite{Mino:2008at,Zimmerman:2011dx} :%
\begin{equation}
\begin{aligned}
Z_{\ell m \omega} & \simeq \hat{D}_{\ell m \omega} \tilde{Z}_{\ell m \omega} \int_{t_0}^{\infty} dt e^{i \omega t - i m \phi(t)} e^{-i (k_{\rm H} + 2i \kappa) x(t)}\,,\\
  \hat{D}_{\ell m \omega} &\coloneqq \frac{1}{2i \omega B_{\ell m \omega}^{\rm in}} (k_{\rm H} + i \kappa) (k_{\rm H} + 2 i \kappa)\,,
\label{spectrum_and_screening}
\end{aligned}
\end{equation}
where $\kappa \coloneqq \sqrt{1- \chi^2}/(2 r_+)$ is the BH surface gravity and particle trajectory is parametrized by $(x(t),\phi(t))$; $\tilde{Z}_{\ell m \omega}$ is a constant factor that depends on the trajectory.
%%%%%
%%%%%
The corresponding time-domain Weyl scalar $r\Psi_{4, \ell m}$ at infinity can be obtained through inverse Fourier transform with respect to the retarded time $u$
\begin{align}
    \left[r\Psi_{4}\right]_{\ell m}(u) &= \int \frac{d\omega}{2\pi}e^{-i\omega u}Z_{\ell m \omega}, \nonumber \\
    &= \text{(QNMs)} + \text{(Direct Waves)} ,
\end{align}
with QNMs from the poles of $\hat{D}_{\ell m \omega}$ in Eq.~\eqref{spectrum_and_screening}\footnote{$B_{\ell m \omega}^{\rm in}$ vanishes at QNM frequencies.}, whereas direct waves from the saddle point of the phase
\begin{align}
    \Phi(t,\omega) \coloneqq -i\omega[u-t+x(t)]-im[\phi(t)-\Omega_H x(t)]+2\kappa x(t). \label{eq:dw_total_phase}
\end{align}
The saddle condition $\partial_t\Phi=\partial_\omega\Phi=0$ yields
\begin{equation}
    u=t-x(t) \,,\quad \omega = \omega_G(t) \coloneqq m\hat\Omega(t) -i\hat g(t)  , \label{direct_wave_frequency}
\end{equation}
with 
\begin{align}
\hat{\Omega} \coloneqq \frac{\beta \Omega_H + \Omega}{1+\beta}\,, \ \hat{g} \coloneqq \frac{2 \beta \kappa}{1+\beta}\,,
\end{align}  
and  local velocity $\beta \coloneqq - dx/dt$ and orbital frequency $\Omega \coloneqq d\phi/dt$.
Here each retarded time $u$ at infinity is mapped to a specific emission time $t$ along the plunging trajectory (FIG.~\ref{fig_directwaves}), reflecting the ``directness'' of the wave emission. Consequently, the instantaneous frequency $\omega_{\rm G}(t)$ is modulated by the plunging motion. The saddle-point contribution  to $r\Psi_{4, \ell m}$ (i.e., direct waves) reads
\begin{align}
    \text{(Direct Waves)} \sim \left(\hat{D}_{\ell m \omega} \tilde{Z}_{\ell m \omega}\right)_{\omega=\omega_G[t_{\rm ret}(u)]}\left[\frac{ e^{-i\int \omega_{\rm G} du}}{1+\beta(t)}\right]_{t=t_{\rm ret}(u)}\,, \label{eq_direct_wave}
\end{align}
where we have used $dt = du/(1+\beta)$ and $t_{\rm ret}$ is the solution of $u = t- x(t)$ for a fixed $u$. See more details in the Supplementary Material (SM).

%%%
%%%
\begin{figure}[t]
\centering
\includegraphics[width=1\linewidth]{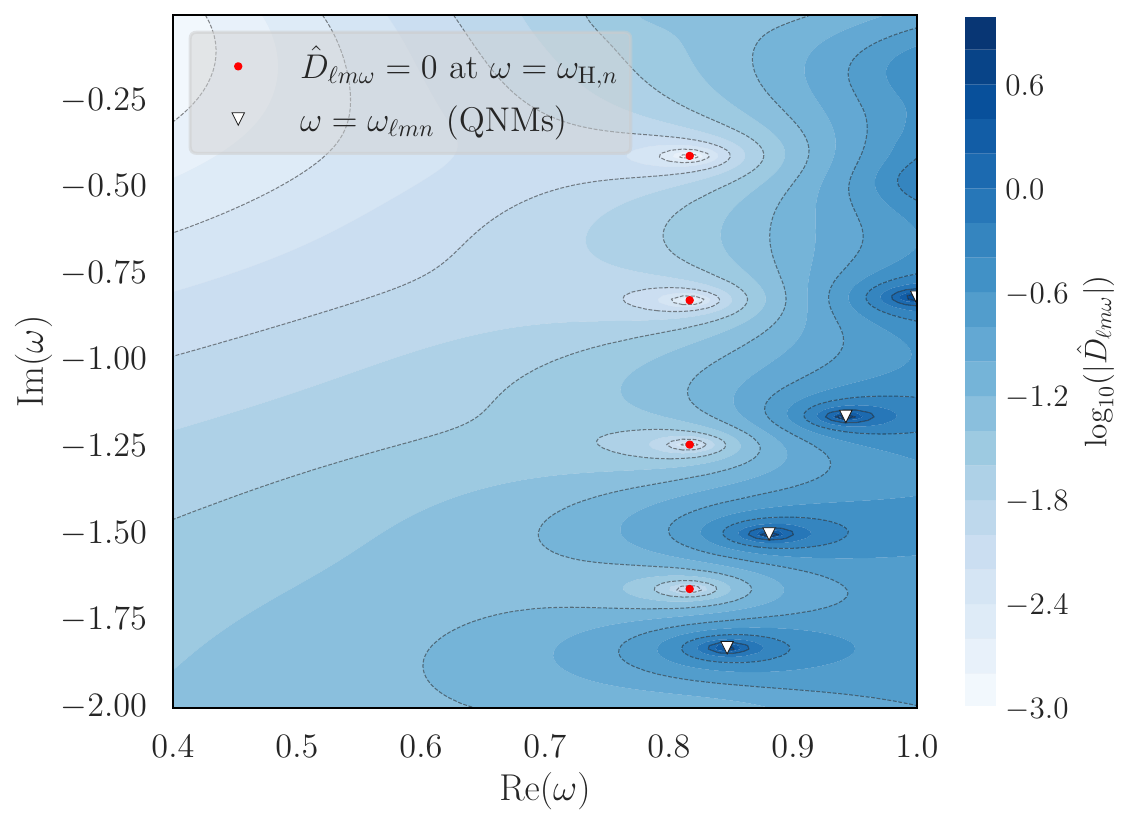}
\caption{
Contour plot of $\log_{10}(|\hat{D}_{\ell m \omega}|)$ in the complex frequency plane for $\chi = 0.7$ and $\ell = m = 2$.
Red dots mark the horizon frequencies $\omega_{{\rm H},n} \coloneqq m \Omega_{\rm H} -in \kappa$ where $\hat{D}_{\ell m \omega_{{\rm H},n}}=0$, and triangles mark QNM frequencies $\omega_{\ell mn}$ where $\hat{D}_{\ell m \omega} \sim (\omega - \omega_{\ell m n})^{-1}$.
}
\label{fig_qnmmap}
\end{figure}
%%%
%%%
At very late times $u\rightarrow + \infty$, the particle approaches the horizon, with $x \to - \infty$, $\Omega \to \Omega_{\rm H}$ and $\beta \to 1$, and the complex frequency $\omega_{\rm G}$ from Eq.~\eqref{direct_wave_frequency} asymptotes to the so-called ``horizon mode'' \cite{Mino:2008at,Zimmerman:2011dx} 
\begin{equation}
\omega_{\rm G} \to m \Omega_{\rm H} - i \kappa \eqqcolon \omega_{\rm H}\,.
\end{equation}
While Mino and Brink \cite{Mino:2008at} suggested that $\omega_{\rm H}$ could leave observable imprints in GWs, Zimmerman and Chen \cite{Zimmerman:2011dx} showed that the factor $\hat{D}_{\ell m\omega}$ in \eqref{spectrum_and_screening} vanishes at $\omega_{\rm H}$, thereby screening this mode from reaching infinity.
Here we further extend this result: the screening is in fact more general:
\begin{equation}
\hat{D}_{\ell m \omega} = 0 \ \text{at} \ \omega = m \Omega_{\rm H} -in \kappa\,,  
\end{equation}
with $n$ an integer, as illustrated in Fig.~\ref{fig_qnmmap}.
However, since $\omega_G(t)$ does not remain fixed at $\omega_H$,
the screening from $\hat D_{\ell m \omega_G}$ does not fully eliminate the direct wave towards infinity. Rather, the observable signal is modulated by $\hat{D}_{\ell m \omega_G(t)}$, which acts as a greybody factor. During the transient plunge stage, the complex frequency $\omega_{\rm G}$ evolves with the instantaneous motion of the particle, as given by Eq.~\eqref{direct_wave_frequency}. In what follows, we explore its time-dependent behavior in simulated waveforms.

%%%%%%%%%%%%%%%%%%%%%%%%%%%%%%%%
\noindent \textbf{\em Numerical simulations.}
%%%%%%%%%%%%%%%%%%%%%%%%%%%%%%%%
A numerical waveform consists of inspiral, merger, and ringdown phases. To reveal the direct wave near merger, one must disentangle it from QNMs, which usually requires accurate mode amplitudes. QNM fits, however, are prone to overfitting, and the number of overtones to include is often uncertain.

We circumvent this issue with QNM filters \cite{Ma:2022wpv,Ma:2023vvr,Ma:2023cwe}, which remove QNMs without relying on fitted amplitudes. In our analysis, we filter out all possible QNMs from the GW strain, $h(u)$, and examine the residual near merger. Details about filters are provided in SM.

\begin{figure}[t]
\centering
\includegraphics[width=1\linewidth]{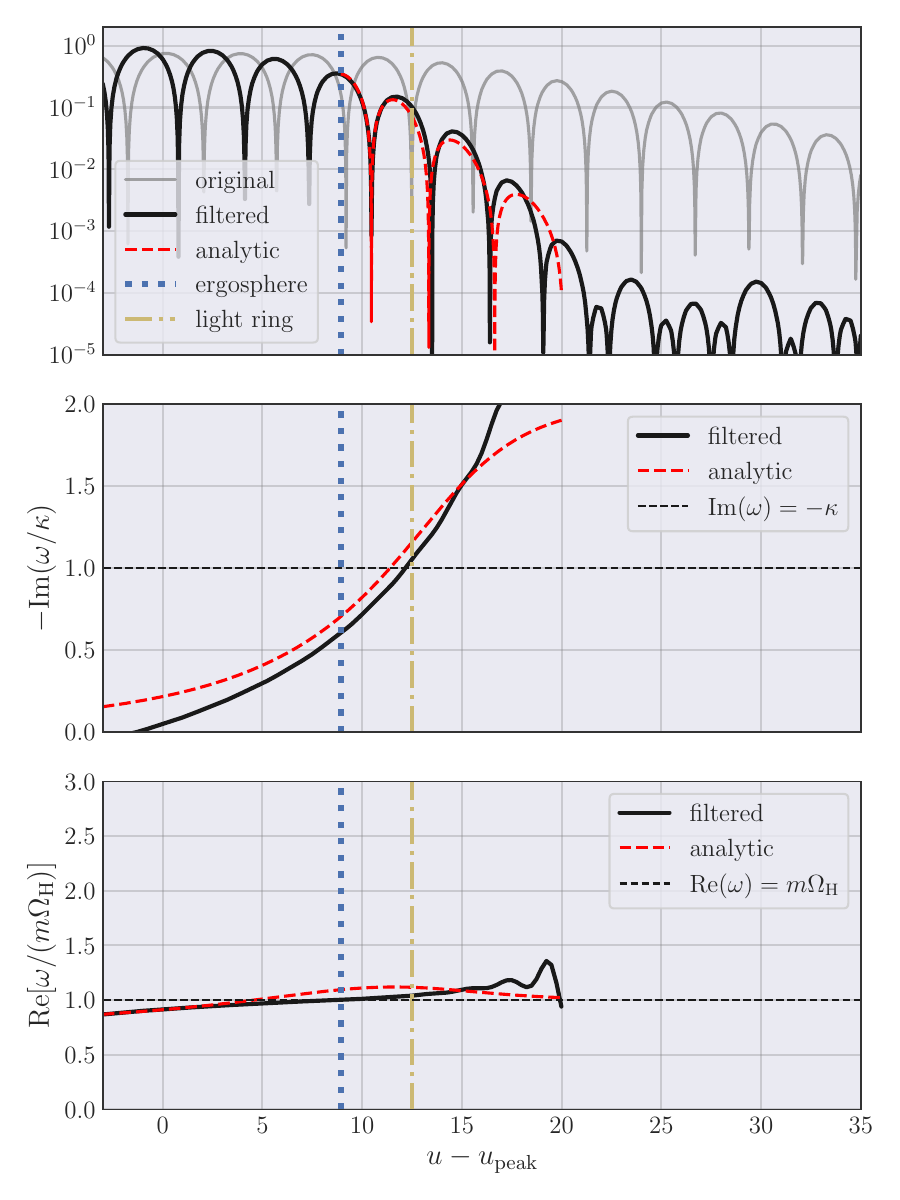}
\caption{
GW strain of an EMRI where the small particle plunges from near the ISCO into a BH with $\chi = 0.8$ (see SM for details).
{\bf (Top)} The $\ell=m=2$ mode after QNM filtering (solid black). 
For reference, the original waveform (solid grey) and a fit with the analytic function describing the direct wave \eqref{eq_direct_wave} (dashed red) are shown.
{\bf (Middle)} Instantaneous decay rate of the filtered waveform (solid black) compared with the evolution of $\text{Im} (\omega_{\rm D})$ from Eq.~\eqref{direct_wave_frequency} (dashed red).
{\bf (Bottom)} Instantaneous frequency of the filtered waveform (solid black) compared with $\text{Re} (\omega_{\rm D})$ (dashed red).
}
\label{fig_EMRI08}
\end{figure}

We first consider EMRIs and compute the quadrupole mode $\ell = m =2$ of the emitted GWs 
\cite{Kojima:1984cj,Saijo:1996iz,Saijo:1998mn,Watarai:2024huy}.
The small particle plunges from near the innermost-stable-circular-orbit (ISCO) into a Kerr BH with $\chi = 0.8$. Geodesic parameters are set based on the Ori-Thorne method \cite{Ori:2000zn}.
The top panel of FIG.~\ref{fig_EMRI08} shows the QNM-filtered strain.
At late times, the filters remove QNMs (gray curve), leaving behind a residual component (black) around merger and early ringdown. This residual oscillates near the superradiant frequency and has a damping rate that rapidly grows and eventually exceeds $\kappa$ (middle and bottom in FIG.~\ref{fig_EMRI08}).
Such behavior appears as the particle passes the ergosphere or the (prograde) light ring (blue dotted and yellow dot-dashed line in FIG.~\ref{fig_EMRI08}, respectively).
The expected strain from Eq.~\eqref{eq_direct_wave} and complex frequency $\omega_{\rm G}$ from Eq.~\eqref{direct_wave_frequency} (red dashed curves) closely follows the simulated results, confirming the residual as the direct wave. Notably, the bottom panel shows that the frequency crosses, rather than asymptotes to, $m\Omega_{\rm H}$, due to the screening factor $\hat{D}_{\ell m \omega}$.
%%%
%%%
%%%
%%%

\begin{figure}[t]
\centering
\includegraphics[width=1\linewidth]{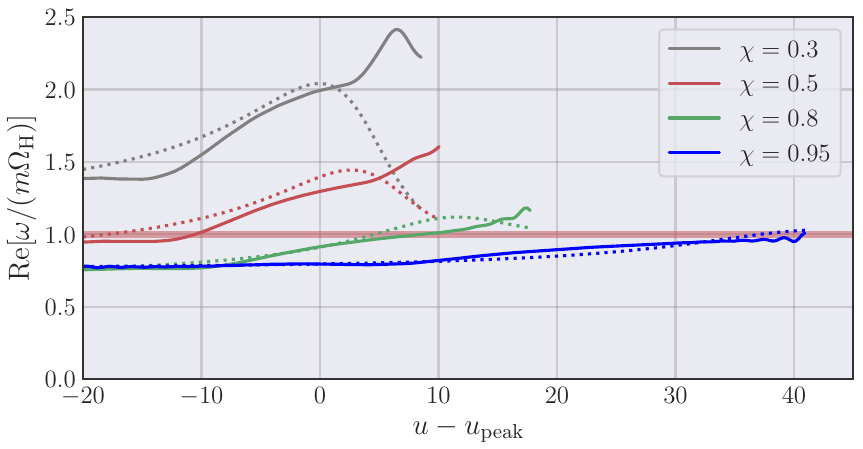}
\caption{
Spin dependence of the extracted instantaneous frequency for EMRIs' direct waves (solid), compared with the corresponding theoretical prediction $\text{Re} (\omega_{\rm D})$ from Eq.~\eqref{direct_wave_frequency} (same color). 
The horizontal thick red line represents the superradiant frequency $m \Omega_{\rm H}$. 
The retarded time $u_{\rm peak}$ is defined at the original strain peak.
}
\label{fig_direct_mode_spin}
\end{figure}

We further examine the spin dependence of the residual.
Figure~\ref{fig_direct_mode_spin} shows its instantaneous frequency between different systems (solid curves), compared to the theoretical expectation of Eq.~\eqref{eq_direct_wave} (dotted curves). For spins $\gtrsim 0.7$, the frequency exhibits a quasi-stable feature near the superradiant value, $\mathrm{Re}(\omega_{\rm D}) \simeq m\Omega_{\rm H}$, which reflects enhanced frame-dragging effects around rapidly rotating horizons: as a particle approaches the horizon equatorially, $d\phi/dt \to \Omega_{\rm H}$ regardless of its orbital constants.
This near-universal frequency $\sim m\Omega_{\rm H}$ can therefore be viewed as a generic feature for high-spin BHs, while deviations from $m\Omega_{\rm H}$ are more pronounced at low spins.

\begin{figure}[t]
\centering
\includegraphics[width=1\linewidth]{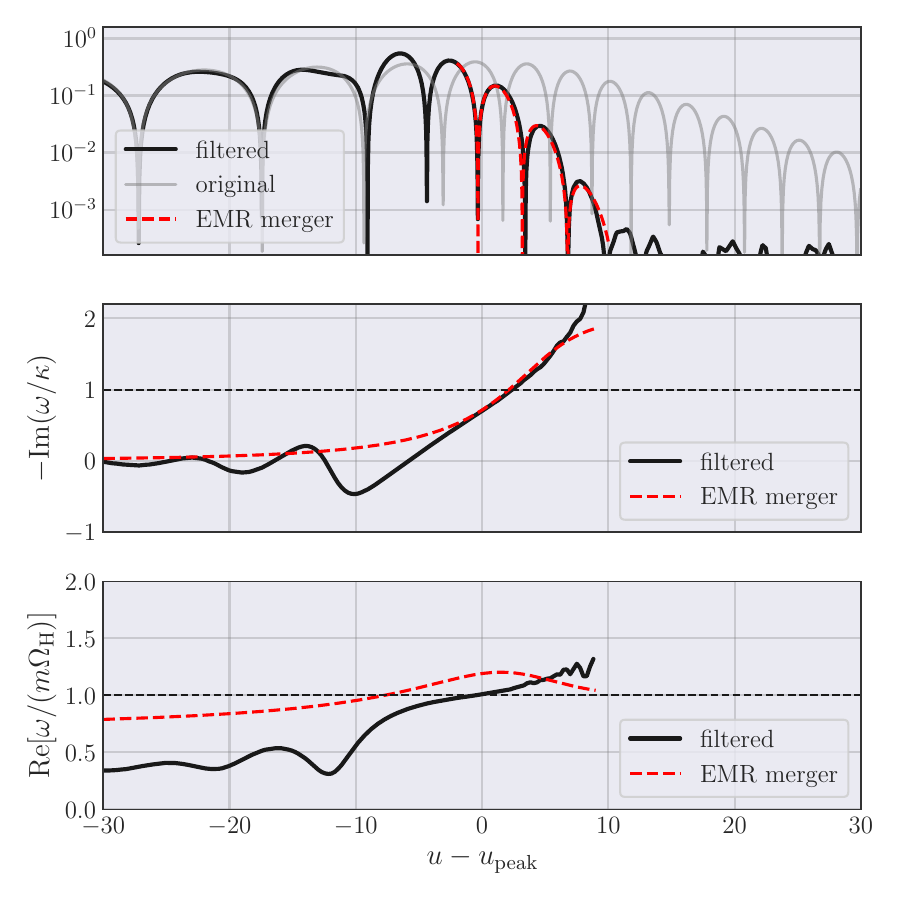}
\caption{
GW strain of a comparable-mass system SXS:BBH:0305.
{\bf (Top)} Real part of the $\ell = m = 2$ mode: QNM-filtered (black) and original (grey).
{\bf (Middle)} Instantaneous decay rate of the filtered waveform normalized by $\kappa$ (black solid), with $\text{Im} (\omega) = - \kappa$ shown as the black dashed line.
{\bf (Bottom)} Instantaneous frequency of the filtered waveform normalized by $m \Omega_{\rm H}$ (black solid), with $\omega = m \Omega_{\rm H}$ as a black dashed line.
The original strain peak is at $u_{\rm peak}$.
For comparison, the time evolution of $\omega_{\rm D}$ (red dashed) from an extreme-mass-ratio (EMR) system is shown in the three panels. 
The particle plunges from the ISCO into a BH with spin $\chi = 0.6921$, matching SXS:BBH:0305.
The SXS strain peak is assumed to correspond to the particle crossing the prograde light ring.
}
\label{fig_direct_mode_SXS}
\end{figure}

Similar features also appear in comparable-mass BBH systems from numerical relativity\footnote{Comparable- and extreme-mass-ratio waveforms are similar under appropriate scaling \cite{Rifat:2019ltp,vandeMeent:2020xgc,Islam:2022laz,Lousto:2022hoq,Planas:2024vnq}.}. We apply QNM filters to the SXS waveforms \cite{Scheel:2025jct}, and find direct waves around the strain peak in most cases. As an example, we consider a GW150914-like system SXS:BBH:0305, and show its original (gray) and filtered (black) strains in the top panel of FIG.~\ref{fig_direct_mode_SXS}. We further compare them with the linear prediction in Eqs.~\eqref{direct_wave_frequency} and \eqref{eq_direct_wave} for an EMRI plunge from the ISCO into a BH with identical remnant parameters (red dashed curves). The agreement is overall good, with small discrepancies likely arising from (i) imperfections in mapping EMRI to comparable-mass systems and/or (ii) nonlinear effects.

%%%
%%%
%%%
%%%

%%%
%%%
%%%
%%%

To summarize, our analysis reveals residual components around the merger of EMRI and SXS waveforms after filtering out QNMs. These direct waves encode information about plunge dynamics and screening effects; and are closely related to the horizon modes and the ``red-shift modes'' discussed in the literature \cite{Mino:2008at,Zimmerman:2011dx,DeAmicis:2025xuh}. In practice, however, such modes are screened by $\hat{D}_{\ell m \omega}$, 
rendering a mode-based description inadequate. Instead, direct waves exhibit time-dependent frequencies and faster damping, with decay rates around $- \text{Im} (\omega_{\rm D}) \sim \mathcal{O}(\kappa)$. 
For rapidly spinning BHs, direct waves have a quasi-stable instantaneous frequency around $m\Omega_{\rm H}$, due to frame dragging (FIG.~\ref{fig_EMRI08} and \ref{fig_direct_mode_spin}).
These features provide novel observables for probing the strongly relativistic regime of (spinning) BHs.

%%%%%%%%%%%%%%%%%%%%%%%%%%%%%%%%
\noindent \textbf{\em Origin of direct waves.}
%%%%%%%%%%%%%%%%%%%%%%%%%%%%%%%%
To numerically identify where direct waves originate, we locally switch off (or ``mask'') the particle source and identify where the dominant component is excited. 
Specifically, we:
(i) mask the source term with
\begin{equation}
{\cal S}_{\ell m \omega} \to {\cal S}_{\ell m \omega} \times W(x_{\rm max}; x)\,,
\end{equation}
where $W$ is a masking function to switch off the source for $x \leq x_{\rm LR} + x_{\rm max}$, with $x_{\rm LR}$ the prograde light ring radius,
\begin{align}
\begin{split}
W\coloneqq \frac{1}{2} + \frac{1}{2} \tanh \left[\frac{x-(x_{\rm LR} + x_{\rm max})}{\sigma} \right] \,,
\end{split}
\end{align}
and $\sigma$ a constant. 
(ii) Compute the strain amplitude with (\ref{teukolsky_R}) and perform the QNM filtering.
(iii) Vary $x_{\rm max}$ and identify a part of the trajectory sourcing direct waves.
To facilitate interpretation, we focus here on a radial plunge from infinity, rather than the ISCO plunge studied earlier (see SM for details).
%
%%% 
\begin{figure}[t]
\centering
\includegraphics[width=1\linewidth]{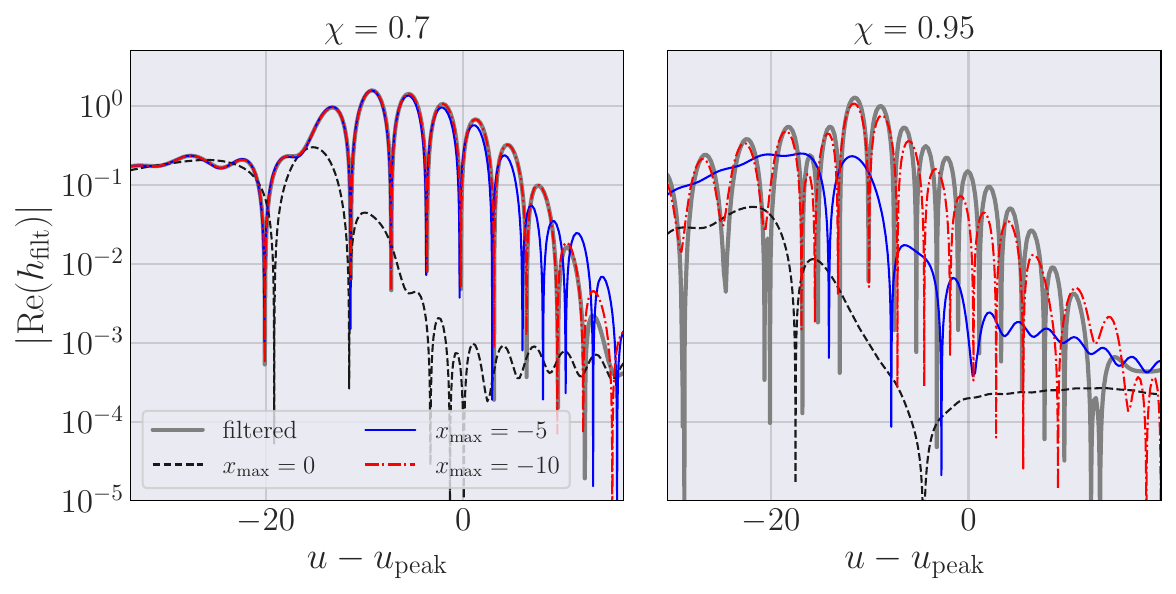}
\caption{
QNM-filtered strains from masked sources plunging into BHs with $\chi = 0.7$ (left) and $\chi = 0.95$ (right).
We consider $x_{\rm max} = 0$ (black dashed), $-5$ (red dot-dashed) and $-10$ (blue thin solid), with $\ell = m =2$ and $\sigma = 0.1$.
Unmasked waveforms are shown for reference (grey thick solid). 
The original strain peak is at $u = u_{\rm peak}$.
See SM for details.
}
\label{fig_switch_source}
\end{figure}
%%%

As shown in FIG.~\ref{fig_switch_source}, masking the source inside the light ring (black dashed curves, $x_{\rm max}=0$) almost completely suppresses direct waves, implying that the components during the merger–ringdown phase are predominantly sourced near the light ring, although a smaller fraction can be emitted prior to the light-ring passage (see Fig.~\ref{fig_EMRI08}).
In contrast, for $x_{\rm max} = -5$ (blue curves), the suppression is stronger for the near-extremal spin ($\chi = 0.95$) than for the moderate case ($\chi = 0.7$).
This behavior is consistent with the gravitational potential barrier having a width of $\sim 2\kappa^{-1}$ in the tortoise coordinate.
The screening function $\hat{D}_{\ell m \omega}$ --- representing the transmissivity of the curvature potential --- efficiently screens direct waves sourced at $x \lesssim - 2\kappa^{-1}$.
Therefore, direct waves are insensitive to masking at $x_{\rm max} \lesssim - 2\kappa^{-1}$, which corresponds to $-5$ for $\chi = 0.7$ and $-8$ for $\chi = 0.95$.

%%%%%%%%%%%%%%%%%%%%%%%%%%%%%%%%
\noindent \textbf{\em Detectability.}
%%%%%%%%%%%%%%%%%%%%%%%%%%%%%%%%
We finally estimate the detectability of direct waves by computing their SNR. Similar to QNM analysis that often assumes a specific starting time, a regular frequency-domain calculation requires smooth time windows to isolate direct waves from earlier signals. However, such windowing often compromises the estimate by introducing contamination from earlier contributions or by obscuring the precise start time of the analysis \cite{Isi:2021iql}. One way to bypass the challenge is to remain in the time domain, where the optimal SNR is given by \cite{Isi:2021iql} 
\begin{align}
    {\rm SNR}=\sqrt{\braket{h|h}}, \label{eq:SNR}
\end{align}
with the inner product defined as
\begin{align}
    \braket{h|h}=\sum_{t_i,t_j>\bar{t}}h_{t_i}C_{ij}^{-1}h_{t_j}.
\end{align}
Here $C_{ij}$ is the autocovariance matrix. As shown in Fig.~\ref{fig_direct_mode_SXS}, the direct wave dominates the late-time portion of the filtered waveform once all QNMs are removed. It is therefore sufficient to insert the filtered waveform into Eq.~\eqref{eq:SNR}, and treat the start time $\bar{t}$ as a free parameter.

\begin{table}
    \centering
    \caption{SNRs for different portions of a GW150914-like system, including the direct wave with various start times, the full inspiral-merger-ringdown signal, and the post-peak signal. Results for LIGO and Virgo are computed at different observational stages, using noise curves from \cite{PSDs}. The O1 curves are extracted directly from the Livingston and Handford data around GW150914.}
    \begin{tabular}{c c c c c c c} \hline\hline
   & & \multicolumn{3}{c}{Direct wave} & \multirow{2}{*}{IMR} & \multirow{2}{*}{Post-peak} \\
   & & $-10M_{t}$ & $-5M_{t}$ & $0M_{t}$  \\ \hline
   \multirow{2}{*}{O1} & LL & 5.6 & 2.0 & 1.1 & 17.2 & 8.9  \\
   & H & 6.2 & 2.2 & 1.2 & 19.6 & 9.2 \\ \hline 
   \multirow{3}{*}{O3} & LL & 13.5 & 4.7 & 2.6 & 38.4 & 17.2 \\
   & H & 10.4 & 3.8 & 2.0 & 31.4 & 16.2 \\
   & V & 3.9 & 1.2 & 0.6 & 14.6 & 5.2 \\ \hline
   \multirow{2}{*}{O4} & L & 19.0 & 7.1 & 3.9 & 54.6 & 24.6 \\
   & V & 11.9 & 4.5 & 2.4 & 34.7 & 15.9 \\ \hline
   \multirow{2}{*}{O5} & A+ & 35.3 & 13.3 & 7.6 & 98.7 & 45.0 \\
   & V & 26.5 & 9.6 & 5.9 & 78.3 & 32.6 \\ \hline
   \multicolumn{2}{c}{CE} & 367.1 & 132.2 & 92.2 & 1530.7 & 404.0 \\ \hline
   \multicolumn{2}{c}{ET} & 251.0  & 81.7& 48.0& 813.1 & 297.0  \\ \hline\hline
     \end{tabular}
     \label{table:SNR}
\end{table}

As a benchmark, we still adopt SXS:BBH:0305, with extrinsic parameters matched to GW150914 \cite{LIGOScientific:2016aoc}. We first consider detector noise extracted from the LIGO Livingston (LL) and Hanford (H) data around the event. As summarized in Table \ref{table:SNR}, the network SNR for the full inspiral-merger-ringdown (IMR) signal and for the post-peak portion is 26.1 and 12.8, respectively --- consistent with those reported in the literature \cite{LIGOScientific:2016aoc,Isi:2019aib}. The corresponding network SNRs of the direct wave are 8.4, 3.0, and 1.6 for start times $\bar{t}=-10M_t,-5M_t$, and $0M_t$. Here we note that the instantaneous frequency of the direct wave at $-5M_t$ is about $98\%$ of the superradiant frequency. 

We also compute SNRs for LIGO and Virgo at different observation stages, using public noise curves \cite{PSDs}, as well as for the Einstein Telescope \cite{ET_website,Hild:2008ng} and Cosmic Explorer \cite{CE_website,Srivastava:2022slt,Evans:2021gyd}. These results are summarized in Table \ref{table:SNR}. Remarkably, for the current three-detector network (O4), direct waves can already reach an SNR of 11.0 at $-5M_t$ and 29.4 at  $-10M_t$, highlighting their potential for near-term detection.

%%%%%%%%%%%%%%%%%%%%%%%%%%%%%%%%
\noindent \textbf{\em Discussion.}
%%%%%%%%%%%%%%%%%%%%%%%%%%%%%%%%
We have identified direct waves in both EMRI and comparable-mass binary waveforms around merger. This additional component during the merger-ringdown phase originates from the instantaneous motion of the companion near the ergosphere and light ring; and its observable signature, partially screened by the gravitational potential barrier, exhibits characteristic modulation and damping. We further show that direct waves can reach detectable SNR with the current LIGO–Virgo–KAGRA network (O4).
These findings mark a concrete step toward shifting the paradigm of BH spectroscopy: from analyzing QNMs alone to uncovering non-modal signals with imminent detectability in current GW events.

The importance of direct waves is multifold.
First, they contribute non-negligibly in the post-peak regime and must therefore be included in QNM decompositions (especially when fitting overtones). This necessity is reflected in the introduction of pseudo-QNMs \cite{Pan:2013rra,Taracchini:2013rva,Babak:2016tgq} in effective-one-body models, and in recent analyses of numerical-relativity waveforms \cite{Giesler:2024hcr}, which show reduced accuracy of QNM fits immediately after the strain peak.
Second, in rapidly spinning BHs, direct waves exhibit quasi-stable frequencies near $\Omega_{\rm H}$, reflecting frame dragging close to the horizon. They may thus provide a direct probe of frame-dragging in the ergosphere.
Finally, the BH greybody factor screens horizon modes and modulates direct waves, offering a novel observational target in GWs.

Our present interpretation of direct waves relies on the BH linear perturbation theory, which may be insufficient for comparable mass mergers, as suggested by the discrepancy in Fig.~\ref{fig_direct_mode_SXS}. While current efforts to model nonlinearity in ringdown focus mainly on quadratic QNMs, our results indicate that direct waves may provide an additional basis for nonlinear effects around merger. This motivates a systematic study of quadratic couplings among direct waves, QNMs, and tails, which will be essential for understanding the transition from the post-Newtonian regime through merger to ringdown, as well as the onset and dynamics of plunges. Such theoretical advances will enable the construction of accurate templates for detecting direct waves in current and future GW observations.

%%%%%%%%%%%%%%%%%%%%%%%%%%%%%%%%%%%%%%%%%%%%%%%%%%%%%%%%%
\noindent \textbf{\em Data availability.}
%%%%%%%%%%%%%%%%%%%%%%%%%%%%%%%%%%%%%%%%%%%%%%%%%%%%%%%%%
The data that support the findings of this article are openly available \cite{oshita_2025_17096498}.

%%%%%%%%%%%%%%%%%%%%%%%%%%%%%%%%%%%%%%%%%%%%%%%%%%%%%%%%%
\noindent \textbf{\em Acknowledgments.}
%%%%%%%%%%%%%%%%%%%%%%%%%%%%%%%%%%%%%%%%%%%%%%%%%%%%%%%%%
%
We are grateful to Aaron Zimmerman for pointing out his earlier work on the screening of the scalar horizon mode by the greybody factor, presented at APS meeting in collaboration with Zach Mark and Y.~C.

N.O. is supported by Japan Society for the Promotion of Science (JSPS) KAKENHI Grant No.~JP23K13111 and by the Hakubi project at Kyoto University.  Y.C.\ is supported by the Brinson Foundation, the Simons Foundation (Award Number 568762), and by NSF Grants PHY-2309211 and PHY-2309231.
H.Y. is supported by the Natural Science Foundation of China (Grant 12573048).
Research at Perimeter Institute is supported in part by the Government of Canada through the Department of Innovation, Science and Economic Development and by the Province of Ontario through the Ministry of Colleges and Universities.

\appendix
\section*{Supplemental Material} 
\subsection{Derivation of Eq.~\eqref{eq_direct_wave}}
The time-domain Weyl scalar $r\Psi_{4, \ell m}$ is given by
\begin{align}
    \left[r\Psi_{4}\right]_{\ell m}(u) &= \iint \frac{d\omega dt}{2\pi}\hat{D}_{\ell m \omega} \tilde{Z}_{\ell m \omega}e^{i\Phi(t,\omega)},
\end{align}
where $\Phi(t,\omega)$ is defined in Eq.~\eqref{eq:dw_total_phase}. Using the steepest-descent method with two variables $t$ and $\omega$, we obtain
\begin{align}
    \iint d\omega dt\hat{D}_{\ell m \omega} \tilde{Z}_{\ell m \omega}e^{i\Phi(t,\omega)} \sim \frac{1}{\sqrt{H}} \hat{D}_{\ell m \omega_*} \tilde{Z}_{\ell m \omega_*}e^{i\Phi(t_*,\omega_*)}.
\end{align}
Here $(t_*,\omega_*)$ is the saddle point and $H$ is the corresponding determinant of the Hessian of $\Phi$ with respect to $(t,\omega)$
\begin{align}
    H=
    \begin{vmatrix}
    \partial_{t}^2 \Phi & \partial_{\omega}\partial_t \Phi \\
    \partial_t\partial_{\omega} \Phi & \partial_{\omega}^2 \Phi
    \end{vmatrix}=(1+\beta)^2.
\end{align}
The phase $\Phi$ at the saddle point reads
\begin{align}
    \Phi(t_*,\omega_*) &= -im[\phi(t_*)-\Omega_H x(t_*)]+2\kappa x(t_*), \notag \\
    &=-i\int^u \omega_G(u^\prime) du^\prime,
\end{align}
where we have used $dt = du/(1+\beta)$.

On the other hand, a Taylor expansion of the trajectory allows one to obtain both the direct wave and the QNMs as pole contributions simultaneously. This procedure is essentially equivalent to the method of steepest descent.
To isolate contributions to the spectrum \eqref{spectrum_and_screening} from different orbital segments, we divide the integral in \eqref{spectrum_and_screening} into pieces $[t_i,t_{i+1}]$ (Fig.~\ref{fig_directwaves}) and decompose the total Weyl scalar $\Psi_{4, \ell m}$ at infinity as $r\Psi_{4,\ell m}=\sum_i\psi_{\ell m[i]}$. Each segment has the form of
\begin{align}
\begin{split}
&\psi_{\ell m [i]} \simeq \int d \omega \frac{-i \hat{D}_{\ell m \omega} \tilde{Z}_{\ell m \omega} S_{\ell m \omega}}{(1+\beta)(\omega - \omega_{\rm G})} \\
&\times \left[ e^{ -i \omega [u-u_i -(1+\beta)\Delta t] -i (1+\beta) \omega_{\rm G} \Delta t}
- e^{ -i \omega (u-u_i)}
\right]\,,
\end{split}
\label{direct_wave_spec}
\end{align}
where $u_i \coloneqq t_i - x(t_i)$.
It has the poles in $\hat{D}_{\ell m \omega}$ at QNM frequencies and in another factor $1/(\omega - \omega_{\rm G})$ leading to the local direct wave.
After inverse-Fourier transforming into the time domain, \eqref{direct_wave_spec} reads
\begin{align}
\begin{split}
&\psi_{\ell m [i]} = {\cal N} \hat{D}_{\ell m \omega_{\rm G}} \frac{e^{ -i \omega_{\rm G} (u-u_i)}}{1 + \beta} + \text{(QNMs)}\,,\\
&\text{for} \ u_i < u < u_i +\Delta u \,,
\end{split}
\end{align}
and vanishes otherwise.
Here ${\cal N}$ is an irrelevant prefactor.

\subsection{Numerical computation and resolution test: Gravitational wave sourced by a plunging particle}
We follow \cite{Kojima:1984cj,Saijo:1996iz,Saijo:1998mn,Watarai:2024huy} to compute gravitational waves (GWs) sourced by a particle plunging into a spinning black hole.
The Sasaki-Nakamura (SN) equation \cite{Sasaki:1981sx}:
\begin{equation}
\left[ \frac{d^2}{dx^2} -{\cal F}_{\ell m \omega} \frac{d}{dx} -{\cal U}_{\ell m \omega} \right] X_{\ell m \omega} = {\cal S}_{\ell m \omega}\,,
\end{equation}
is solved in the frequency domain using the fourth Runge-Kutta method. Here $X_{\ell m \omega} (x)$ is the SN perturbation variable. The explicit forms of ${\cal F}_{\ell m \omega}$ and ${\cal U}_{\ell m \omega}$ are given in \cite{Sasaki:1981sx}, while the source term ${\cal S}_{\ell m \omega}$ can be found in \cite{Kojima:1984cj, Saijo:1996iz,Watarai:2024huy}.

The geodesic motion of the particle is parameterized in Boyer-Lindquist coordinates as $\{t = t(\tau), r = r(\tau), \theta = \theta (\tau), \phi = \phi (\tau)\}$, where $\tau$ is the particle's proper time.
Throughout, we restrict to equatorial trajectories $(\theta = \pi/2)$, and solve the motion adiabatically by fixing the particle's energy $E$ and orbital angular momentum $L_z$.
The geodesic equations for a plunging orbit $dr/d\tau < 0$ on the equatorial plane are
\begin{align}
r^2 \frac{dt}{d\tau} &= -a (a E - L_z) + \frac{r^2 + a^2}{\Delta(r)} P(r)\,,\\
r^2 \frac{dr}{d\tau} &= -\sqrt{Q(r)}\,,\\
r^2 \frac{d\phi}{d\tau} &= - (a E-L_z) + \frac{a}{\Delta(r)} P(r)\,,
\end{align}
where $P(r) \coloneqq E (r^2+a^2) -aL_z$ and $Q(r) \coloneqq P^2(r) -\Delta(r) \left[ r^2 +(L_z -aE)^2 \right]$.

\begin{figure}[t]
\centering
\includegraphics[width=0.48\linewidth]{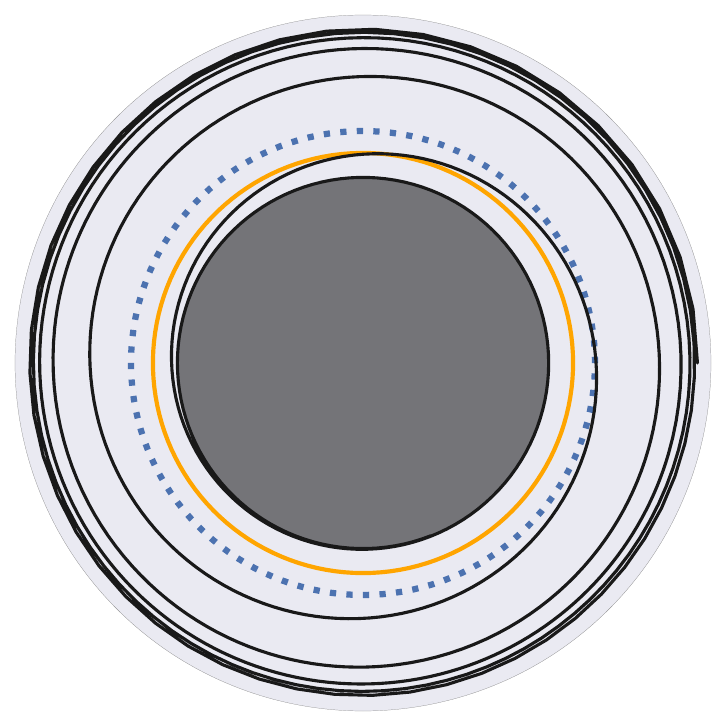}
\includegraphics[width=0.48\linewidth]{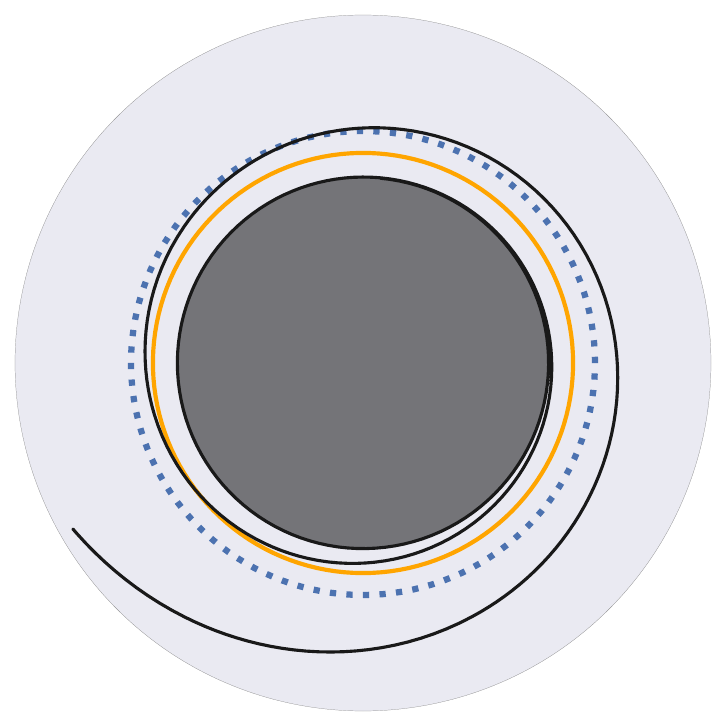}
\caption{
Trajectory of a particle plunging equatorially into a Kerr BH with  $\chi = 0.8$, from the ISCO (left) and from the far region (right)  
As a reference, the ergosphere (blue dotted line) and the prograde light ring (yellow solid line) are shown.
}
\label{fig_orbits}
\end{figure}

We considered two types of geodesic motions: (i) a plunge from the innermost stable circular orbit (ISCO) (FIG.~\ref{fig_EMRI08} and \ref{fig_direct_mode_spin}) and (ii) a radial plunge from infinity (FIG.~\ref{fig_switch_source}).
The corresponding particle trajectories are shown in FIG.~\ref{fig_orbits}.
%%% 
%%%

For case (i), we use the Ori-Thorne method to introduce realistic small deviations $\Delta E \coloneqq E- E_{\rm ISCO}$ and $\Delta L_z \coloneqq L_z- L_{\rm ISCO}$ from the corresponding ISCO values, to avoid the scenario where the particle circulates indefinitely on the ISCO without plunging. The deviations depend on the mass ratio $q$ \cite{Ori:2000zn} (see also, \cite{Watarai:2024huy}); throughout we take $q = 10^{-5}$.

For case (ii), the particle energy is set to its rest mass, $E = \mu$, and the orbital angular momentum to $L_z = 0.99 \times L_{z {\rm (c)}}$, where $L_z = L_{z {\rm (c)}}$ is the critical value above which the particle does not plunge into the black hole. 

In our numerical calculations, the SN equation is solved at fixed frequencies $\omega$, with frequency resolution $\Delta \omega$
\begin{align}
\Delta \omega = (\omega_{I+1} - \omega_{I})/N_{\rm f},
\end{align}
in the range of $\omega_I \leq \omega < \omega_{I+1}$, where $\{\omega_{0}, \omega_1, \omega_2, \omega_3\} = \{0.001, 0.1, 1, 2 \}$ and $N_{\rm f} = 144$ is the number of available CPU cores. Spectral amplitude in the negative-frequency region is also computed in the same manner.
%%% 
\begin{figure}[t]
\centering
\includegraphics[width=1\linewidth]{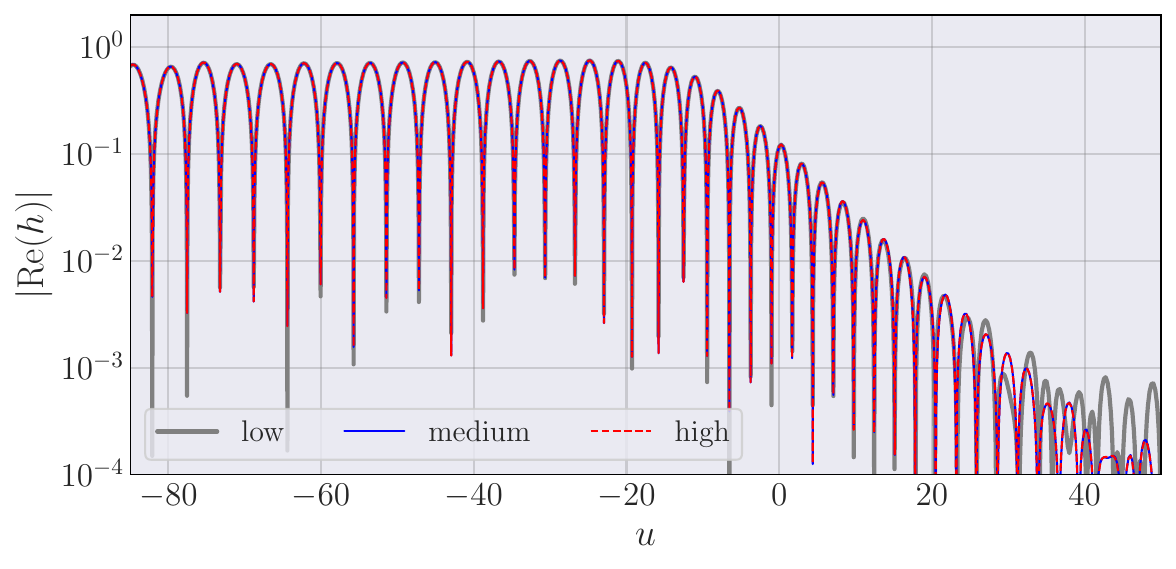}
\caption{
The $\ell = m = 2$  mode of GW strain sourced by a particle plunging from the ISCO into a Kerr BH with $\chi = 0.8$. Numerical resolutions are $N_x = 10^4$ (low), $2 \times 10^4$ (medium), and $3 \times 10^4$ (high), respectively.
}
\label{fig_resolution}
\end{figure}
%%%
The spatial resolution in the tortoise coordinate is set as
\begin{equation}
\Delta x = (x_{\rm out} - x_{\rm inn})/N_x\,.
\label{delta_x}
\end{equation}
For $\chi = 0.8$ and case (i), we choose
$x_{\rm inn} = -20$, $N_x = 3 \times 10^{4}$, and $x_{\rm out} \coloneqq x (r = r_{\rm ISCO} - q^{2/5}R_o)$, where  the explicit form of $R_o$ is given in \cite{Watarai:2024huy}. For $\chi = 0.7$ and case (ii), we select $x_{\rm out} = 200$, $x_{\rm inn} = -20$ and $N_x = 3 \times 10^4$.
We have checked that the radial resolution is sufficient for waveform convergence (FIG.~\ref{fig_resolution}). 
The frequency and spatial ranges are adjusted depending on the black hole spin and the particle trajectory.

\subsection{QNM filtering}
We employ QNM rational filters\cite{Ma:2022wpv,Ma:2023vvr,Ma:2023cwe} to remove QNMs from the strain, $h(u)$, through
\begin{align}
h_{\rm filt} (u) &\coloneqq \frac{1}{2 \pi} \int d \omega \ \tilde{h}(\omega) F_{\rm tot} e^{-i \omega u}\,,
\end{align}
where $\tilde{h} (\omega) \coloneqq \int d u h (u) e^{i \omega u}$ is the Fourier transform of $h(u)$, and $F_{\rm tot}$ is the total filter consisting of multiple prograde and retrograde QNMs:
\begin{equation}
F_{\rm tot} \coloneqq \left(\prod_{n=0}^{N_{\rm p}} F_{\ell m n}^{\rm (P)} \right) \times \left(\prod_{n=0}^{N_{\rm r}} F_{\ell m n}^{\rm (R)} \right)\,,
\end{equation}
with
\begin{equation}
    \begin{aligned}
    F_{\ell m n}^{\rm (P)} &\coloneqq \frac{\omega -  \omega_{\ell m n}}{\omega - \omega_{\ell m n}^*}\,, \\
    F_{\ell m n}^{\rm (R)} &\coloneqq \frac{\omega + \omega_{\ell -m n}^*}{\omega + \omega_{\ell -m n}}\,.
    \end{aligned}
\end{equation}
%%% 
\begin{figure}[b]
\centering
\includegraphics[width=1\linewidth]{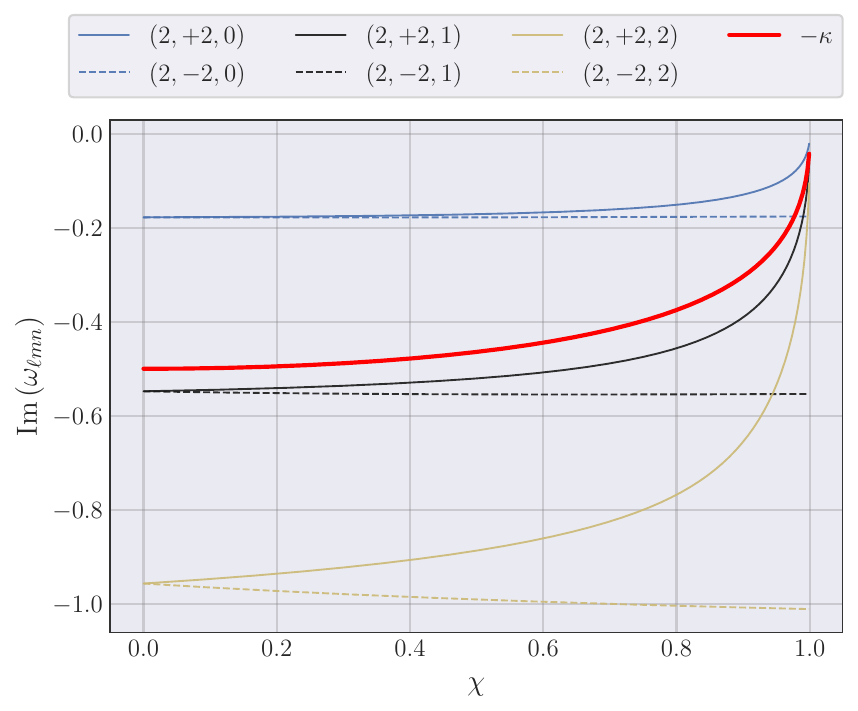}
\caption{
Comparison between the surface gravity, $- \kappa$, (red thick line) and the decay rates of various QNMs for $|s| = 2$ and $(\ell, m ,n= 0,1,2)$. 
Prograde and retrograde QNMs are plotted with thin solid and thin dashed lines, respectively.
QNM values are obtained from \cite{grit_ringdown_data}.
}
\label{fig_qnm_kappa}
\end{figure}
%%%
Here $N_{\rm p}$ and $N_{\rm r}$ are the number of included prograde and retrograde overtones.
We set $N_{\rm p} = 7$ and $N_{\rm r} = 2$ for SXS:BBH:0305 in FIG.~\ref{fig_direct_mode_SXS}; and $N_{\rm p} \geq 7$ and $N_{\rm r} \geq 2$ for extreme-mass-ratio plunges. The choice is sufficient to reveal direct waves, since they decay slower than the second and higher overtones: $\text{Im} (\omega_{\ell |m| 2}) < -\kappa \sim \text{Im} (\omega_{\ell m 1})$, at least for $\ell = m =2$ and $|s| = 2$ (see FIG.~\ref{fig_qnm_kappa}).
For SXS:BBH:0305, we also filter the spherical-spheroidal mixing mode $(3,2,0)$.

\end{document}